\documentclass[fleqn,twoside]{article}
\usepackage[headings]{espcrc2}
\readRCS
$Id: andrei_dorokhov_v04.tex,v 0.2 2005/05/09 10:41:11 adorokhov Exp $
\usepackage{graphicx}
\usepackage{epsfig}
\usepackage[figuresright]{rotating}
\DeclareSymbolFont{greek}{U}{eur}{m}{n}
\DeclareMathSymbol{\upalpha}{\mathord}{greek}{"0B}
\DeclareMathSymbol{\upbeta}{\mathord}{greek}{"0C}
\DeclareMathSymbol{\upgamma}{\mathord}{greek}{"0D}
\DeclareMathSymbol{\updelta}{\mathord}{greek}{"0E}
\DeclareMathSymbol{\upmu}{\mathord}{greek}{"16}
\newcommand{\um}{\ensuremath{\upmu}m} 
 
\newcommand{\C} {\ensuremath{^{\circ} \mathrm{C}}} 
\newcommand{\Deg} {\ensuremath{^{\circ}}} 
\newcommand{\Neq} {\ensuremath{\mathrm{n_{eq}/{cm^2}}}}
\newcommand{\E} [1] {\ensuremath{\times 10^{#1}\,}}

\newcommand{\AmS}{{\protect\the\textfont2
  A\kern-.1667em\lower.5ex\hbox{M}\kern-.125emS}}

\title{Extraction of electric field in heavily irradiated silicon pixel sensors}

\author{
A.~Dorokhov
\address[Zurich]{Physik Institut der Universit\"at Z\"urich-Irchel, 8057 Z\"urich, Switzerland}%
\address[PSI]{Paul Scherrer Institut, 5232 Villigen, Switzerland}%
        \thanks{Corresponding author.~Institut de Recherches Subatomiques, 23 rue du loess, BP28, F67037 Strasbourg. 
	{\it E-mail address:} Andrei.Dorokhov@IReS.in2p3.fr},
Y.~Allkofer\addressmark[Zurich],
C.~Amsler\addressmark[Zurich],
D.~Bortoletto\address[Purdue]{Purdue University, Task G, West Lafayette, IN 47907, USA},
V.~Chiochia\addressmark[Zurich],
L.~Cremaldi\address[Miss]{Mississippi State Univ., Department of Physics and Astronomy, MS 39762, USA},
S.~Cucciarelli\address[Basel]{Institut f\"ur Physik der Universit\"at Basel, 4056 Basel, Switzerland},
C.~H\"ormann\addressmark[Zurich]\addressmark[PSI],
D.~Kim\address[JHU]{Johns Hopkins University, Baltimore, MD 21218, USA},
M.~Konecki\addressmark[Basel],
D.~Kotlinski\addressmark[PSI],
K.~Prokofiev\addressmark[Zurich]\addressmark[PSI],
C.~Regenfus\addressmark[Zurich],
T.~Rohe\addressmark[PSI],
D.~Sanders\addressmark[Miss],
S.~Son\addressmark[Purdue],
T.~Speer\addressmark[Zurich],
M.~Swartz\addressmark[JHU]
}

\runtitle{Extraction of electric field in heavily irradiated silicon pixel sensors}
\runauthor{A. Dorokhov}
\begin{document}
\begin{abstract}
A new method for the extraction of the electric field in the bulk of heavily irradiated silicon pixel 
sensors is presented. 
It is based on the measurement of the Lorentz deflection 
and mobility of electrons as a function of depth.
The measurements were made at the CERN H2 beam line, 
with the beam at a shallow angle with respect to the pixel sensor surface.
The extracted electric field is used to simulate the
charge collection and the Lorentz deflection in the pixel sensor.
The simulated charge collection and the Lorentz deflection is in good agreement 
with the measurements both for non-irradiated and irradiated up to $10^{15}$ \Neq
sensors. 
\vskip \baselineskip \noindent PACS: 29.40.Gx; 29.40.Wk; 61.80.-x

\vskip \baselineskip \noindent Key words: Electric field; Radiation hardness; Lorentz angle; Charge collection; Silicon; Pixel; CMS;

\end{abstract}


\maketitle
\section{Introduction}

The properties of the silicon sensors designed for the CMS pixel detector \cite{CMS:TDR}
will change during the LHC operation. The innermost barrel layer of the CMS pixel detector 
is expected to be exposed to a fluence\footnote{All particle fluences are normalized to 
1 MeV neutrons (${\rm n_{eq}}/\mbox{cm}^2$).} of ~$3 \times 10^{14}~{\rm n_{eq}}/\mbox{cm}^2$ per year
at full luminosity. The irradiation dose will be a few times larger in the case of the LHC luminosity upgrade.
The silicon sensors behavior will be determined by the radiation damage,
which changes the electric field in the silicon bulk and introduces charge trapping.
This will lead to a reduction of the collected charge \cite{trohe_florence04}. 
The pixel detector will operate in a 4 T magnetic field and charge carriers will be deflected by the Lorentz force,
which enhances charge sharing between pixels and improves the spatial resolution. 
However, the bias voltage will be increased because of irradiation and the Lorentz deflection
will be reduced \cite{adorokhov_firenze03}.
The spatial resolution depends on the charge collection, track position,
signal, noise and the Lorentz angle, and is degraded by irradiation.
Here we present measurements of charge collection and Lorentz deflection 
as a function of depth in the silicon bulk for heavily irradiated pixel sensors.
A new method for the extraction of the electric field in the silicon bulk is
proposed and validated with a simple simulation.

\section{Sensors and the measurement technique}
\begin{figure}[htb]
\begin{center}
\epsfig{file=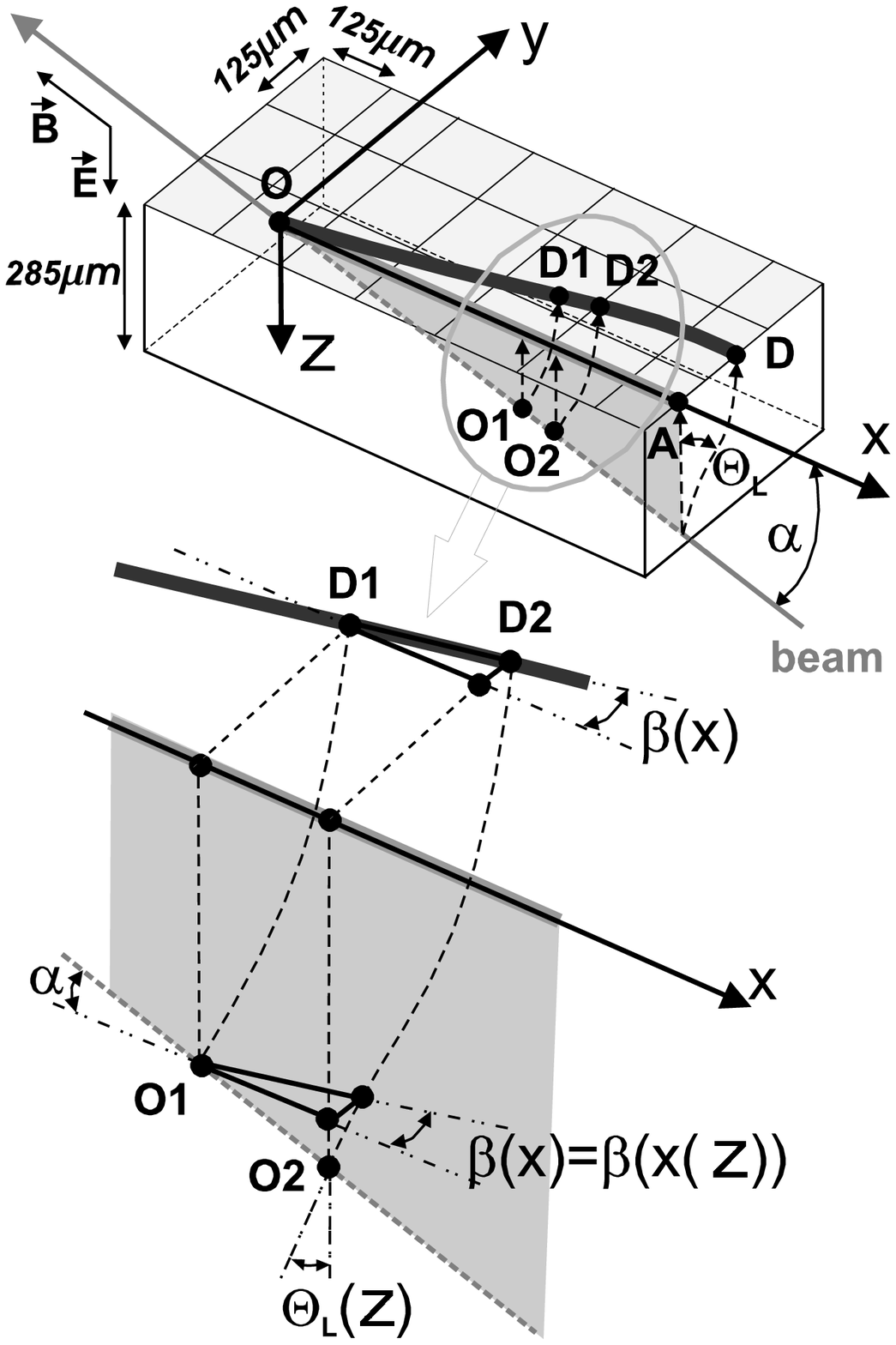,width=0.45\linewidth, bb=174 235 390 697}
\end{center}
\caption{The deflection measurement technique.}
\label{grazingangle_como.eps}
\end{figure}

The sensors under study were designed for the CMS pixel detector and based on  
the ``n-on-n'' concept~\cite{Rohe_IEEE}. 
The bulk material is diffusively-oxygenated float zone (DOFZ) n-type silicon of  
$\rm{\langle 111 \rangle }$ orientation and a resistivity 2-5$\rm{\: k\Omega \: cm}$.
The pixels are formed by p-spray isolated n$^+$-implants, while the p-n junction is formed by
a large p$^+$-implant on the back side.
The thickness of the sensor is 285 \um\ and the pixel size is $\rm{125\times 125}$ \um$^2$.
The sensors were irradiated at the CERN SPS with 24 GeV protons at room temperature
without applying bias voltage and then stored at -20\C.
The tests were carried out at the CERN H2 beam line with 150-225~GeV pions.
The beam entered the pixel plane at a shallow angle $\rm{\alpha=}$15\Deg and the 3~T magnetic field was 
parallel to the beam (see Fig. \ref{grazingangle_como.eps}). 
The relationship between the $z$ position of the created charge carriers 
and the corresponding arrival position at the pixel plane along the $x$ axis is given by $z=x\tan{\alpha}$.
Since the charge is always integrated in the pixel area, the
smaller angle $\alpha$ is used the more precise location of charge 
carriers origin along the $z$ axis is probed.
The position of the beam exit point ``O'' was reconstructed in the 
pixel coordinates system using the beam telescope~\cite{camsler_beam_telescope}.
The beam telescope consisted of four modules, each 
containing two silicon strip sensors measuring  the $x$ and $y$ coordinates.
The strips had a pitch of 25 \um, readout pitch of 50 \um\ and the spatial resolution of each plane was about 1\um.
The pixel sensors were bump-bonded to the $\rm{PSI30/AC30}$ chip~\cite{dmeer_psi30},
which read out all signals from the $\rm{22 \times 32}$ pixel matrix. 
The pixel sensor was cooled by Peltier elements down to -10\C.
Both pixel and beam telescope signals were digitized using VME-based ADC modules
controlled by a DAQ software written in LabView and LabWindows/CVI~(National Instruments).
The trigger was provided by a PIN diode of size $3\times 6$ mm$^2$ placed between
the beam telescope planes before the pixel detector. \\
The electrons and holes produced by 
particles crossing the pixel sensor drift toward the electrodes.
In absence of magnetic field the electrons 
are collected along the segment $\rm{\overline{OA}}$ (see Fig. \ref{grazingangle_como.eps}). 
The holes move to the opposite direction and, together with 
the electrons, induce the net current on the pixels situated along $\rm{\overline{OA}}$. 
In presence of a magnetic field charge carriers are deflected by the
Lorentz force and the resulting current is induced on the pixels along the segment $\rm{\overline{OD}}$. 
This measurement technique was developed in~\cite{Henrich_grazing_angle} and used to measure 
the averaged Lorentz angle, $\Theta_L$, by fitting the deflection $\rm{\overline{OD}}$ with a straight line.
As we will see in section~\ref{Charge_collection_and_the_Lorentz_deflection}, 
the segment $\rm{\overline{OD}}$ is curved, because the
Lorentz angle depends on the electric field, which changes over the depth. 
However, the experimental technique described in ~\cite{Henrich_grazing_angle}
can be applied for measuring the Lorentz angle as a function of depth in the sensor bulk.
\begin{figure}[htb]
\begin{center}\epsfig{file=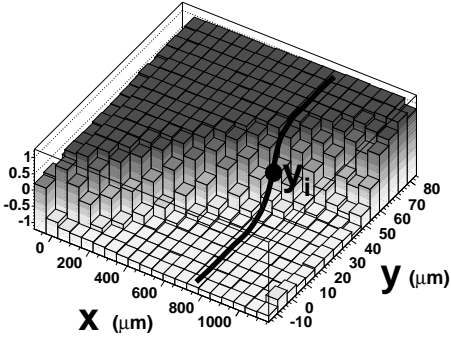,width=0.55\linewidth,bb=100 140 450 430, angle=0,silent=}\end{center}
\caption{Asymmetry as a function of position in the $xy$ plane for a non-irradiated sensor.}
\label{asymmetry.eps}
\end{figure}
The bottom part of Fig.~\ref{grazingangle_como.eps} shows the definition of 
$\beta(x)$ for an infinitely small section of the segment $\rm{\overline{OD}}$.
Knowing the beam incident angle $\alpha$=15\Deg\ and the deflection angle $\beta(x)$
the Lorentz angle at a certain depth $\Theta_{L}(z)$ is given by 
\begin{equation}
\tan{\Theta_L(z)}={{\tan{\beta(x(z))} \over \tan{\alpha}}}={\tan{\beta(x)} / \tan{\alpha}} .
\end{equation}
The angle $\alpha $ is known, therefore, the Lorentz angle is calculated in each point 
in depth with the tangent to the segment $\rm{\overline{OD}}$, i.e. with the $\tan{\beta(x)}$. 
The geometrical position of the segment $\rm{\overline{OD}}$ in the $xy$ coordinates plane 
can be determined from the signal asymmetry of two neighboring pixels.
The asymmetry at the $(x,y)$ position is defined as
\begin{equation}
A(x,y) = \frac{(Q_{x, y-p/2} - Q_{x, y+p/2})}{  (Q_{x, y-p/2}+Q_{x, y+p/2})}, 
\end{equation}
where $Q_{x,y-p/2}$ and $Q_{x,y+p/2}$ is the charge collected in pixel,
whose center is located at $(x,y-p/2)$ and $(x,y+p/2)$, respectively,
and $p=$125 \um\ is the pixel size.
The asymmetry averaged over all events in each $(x,y)$ bin is shown in Fig.~\ref{asymmetry.eps}.
The asymmetry plot was divided into slices along the $x$ axis.
The charge spread is approximated with the Gaussian function, 
therefore the asymmetry in the $i$-th slice located at $x_i$ (e.g. represented by the solid line in Fig.~\ref{asymmetry.eps}) 
was fitted with the standard normal cumulative distribution function of $y$
\begin{equation}
A(x=x_i,y)=c\times {{ {\sqrt{2 \over \pi}}} \int_{-\infty}^{(y-y_i)/s}{e^{-t^2/2 }dt}   - c },
\end{equation}
where the parameter $y_i$ corresponds to the zero asymmetry position along the $y$ coordinate 
for $i$-th slice (see Fig.~\ref{asymmetry.eps}),
$c$ and $s$ are the constant and spread parameters of the fit.
The set of points ($y_i$,$x_i$) determines the segment $\rm{\overline{OD}}$.
%
The $\tan{\beta}$ is determined from the derivative $dy/dx$. 
%
%
The slight rotation (the line $\rm{\overline{OA}}$
can be rotated with respect to the pixel row) of the sensor in the $xy$ plane was subtracted 
using the data without magnetic field.
Each point $x_{i}$ corresponds to a certain depth via the relation $z_i=x_i\tan{\alpha}$ 
and the deflection $y_i$ can be expressed as a function of depth.
The measured points were fitted with a 5-th order polynomial function (see Fig.~\ref{la.eps}).

\section{Charge collection and the Lorentz deflection}
\label{Charge_collection_and_the_Lorentz_deflection}
\begin{figure}[htb]
\begin{center}\epsfig{file=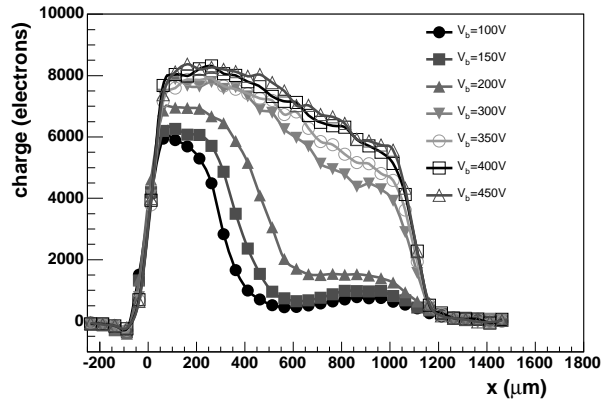,width=\linewidth,bb=20 90 510 316,angle=0,silent=}\end{center}
\caption{Average collected charge as a function of the distance to the exit point 
for a sensor irradiated at 6.7\E{14}\Neq\ for different bias voltages.}
\label{prof5.eps}
\end{figure}

\begin{figure}[htb]
\begin{center}\epsfig{file=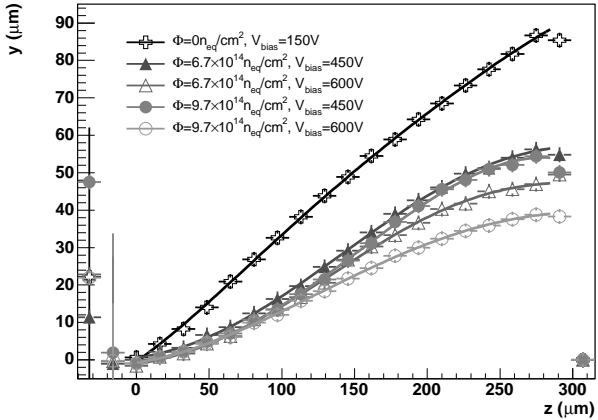,width=\linewidth,bb=20 90 490 316,angle=0,silent=}\end{center}
\caption{Lorentz deflection of the charge in a 3 T magnetic field as a function of the depth.}
\label{la.eps}
\end{figure}

Measurements without magnetic field were performed 
to determine the signal distribution along the 
segment $\rm{\overline{OA}}$ see Fig.~\ref{grazingangle_como.eps}.
Assuming the averaged  energy loss along the particle track to be uniform,
the average signal in a pixel along $\rm{\overline{OA}}$ 
is proportional to the charge collection efficiency 
originating at a certain depth in the silicon bulk.
The average charge collected by a single pixel as a
function of the pixel position along $\rm{\overline{OA}}$ 
is shown in Fig.~\ref{prof5.eps} for a sensor irradiated at 6.7\E{14}\Neq.
One can see, that even at low bias voltage (100-200V) some charge is collected 
from the p$^+$ side and the charge has a minimum 
in the middle of the sensor thickness. 
Most of the charge, however, comes from the region close to the pixel implant.
This behavior can be explained by a non-linear electric field and by trapping of charge carriers. 
If the sensor is operated in a magnetic field, 
the charge carriers are deflected by the Lorentz angle $\Theta_L$.
The signal is induced on the pixels along the segment $\rm{\overline{OD}}$ (see Fig. \ref{grazingangle_como.eps}).
Since the electric field in irradiated sensor is not linear,
the segment between the points ``O'' and ``D'' is curved, 
and in each point its tangent determines the deflection angle $\beta$. 
The deflection of collected charge
along the $y$ axis as a function of the depth in the silicon bulk
is shown in Fig.\ref{la.eps} for different fluences and bias voltages.
It must be noticed, that at the edges of the silicon bulk ($\rm{z<17}$\um\ or $\rm{z>268}$\um) 
the measured deflection has high systematic uncertainties (not shown in Fig.~\ref{la.eps})
due to the geometrical distortions of the electric field lines and incorrect reconstruction of the 
deflection curve $\rm{\overline{OD}}$.
The errors due to the particles multiple scattering were eliminated by 
selecting only straight tracks reconstructed by the beam telescope.

\section{Electric field measurements}
\subsection{Electric field strength across the silicon bulk}
In case of small Lorentz angles the low magnetic field approximation can be used 
to describe the charge carriers motion in crossed electric and magnetic fields.
This approximation physically corresponds to the situation where charge carriers travel
only for a small arc of the circular orbit before scattering moves them from this orbit into another. 
The general expression of the current density
in presence of crossed electric and magnetic fields can be found in~\cite{bube}.
The Lorentz angle can be obtained from the direction of the current density as
\begin{equation}
{ { \tan{\Theta_{L}} } = { r_{h}   }} \mu B ,
\label{lor_angle}
\end{equation}
where $r_{h}$ is the Hall factor, $B$ is the magnetic field and $\mu$ is the drift mobility. 
The charge carriers mobility is determined by the lattice and impurity scattering.
At the temperatures around 300 K and for the impurity concentration
up to 10$^{18}$ cm$^{-3}$ the major contribution to the mobility is from lattice scattering.
For the impurity concentration up to 10$^{15}$ cm$^{-3}$ the impurity scattering contribution
to the total mobility is below one percent and the Hall factor 
changes less than one percent in the range of 
impurity concentration from 10$^{13}$ cm$^{-3}$ to 10$^{15}$ cm$^{-3}$~\cite{norton}. 
Assuming that the irradiation induced defects act as impurity atoms, 
the influence of irradiation on the mobility and Hall factor can be neglected
for all irradiated sensors used for the tests.
Therefore the measured Lorentz angle can be used to calculate the mobility
using the Eq.~\ref{lor_angle}.
Most of the signal is due to the electrons contribution due to their
shorter collection time and due to specific shape of the effective potential
which is confirmed by the simulation (see Fig.~\ref{sim_cl_h_mag.eps} and Fig.~\ref{sim_cl_e_mag.eps}). 
Moreover, despite the fact that holes are also deflected during their drift, they will arrive
to the backside which has the same effective potential used to calculate the induced
current on pixels using the Ramo-Shockley theorem~\cite{shockley-ramo}. 
Therefore the signal induced by holes will be on the line $\rm{\overline{OA}}$
even in presence of magnetic field and will not disturb the measured line $\rm{\overline{OD}}$
which is only due to the collected electrons.
Using the measured $\Theta_{L}(z)$ ($\tan{\Theta_L}$ is a derivative of the deflection shown in Fig.~\ref{la.eps})
and using Eq.~\ref{lor_angle}, the electron mobility as a function of the depth is given by
\begin{equation}
\mu_{e}(z) = { { \tan{\Theta_{L}}(z) }\over { r_{h}  B_{x} }},
\end{equation}
where $r_{h}=1.15$ is the Hall factor for the electrons and $B_{x}=(3\cos(15\Deg))$ T 
is the projection of the magnetic field along the $x$ axis.
\begin{figure}[htb]
\begin{center}\epsfig{file=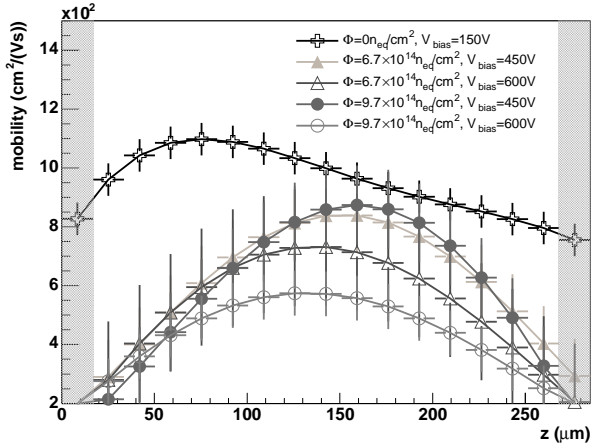,width=\linewidth,bb=30 90 490 325,angle=0,silent=}\end{center}
\caption{Measured electrons mobility as a function of depth for different fluences and bias voltages.
The shaded regions correspond to the depth values where the mobility has large systematic uncertainties.}
\label{mob.eps}
\end{figure}
The measured electron mobility as a function of the sensor depth is shown in Fig.\ref{mob.eps} 
for different fluences and bias voltages. 
\par
\begin{figure}[htb]
\begin{center}\epsfig{file=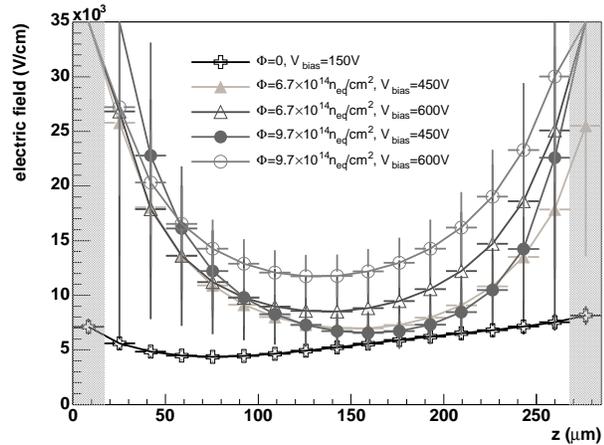,width=\linewidth,bb=30 90 490 320,angle=0,silent=}\end{center}
\caption{Extracted electric field as a function of depth for the non-irradiated and irradiated
silicon sensors at different bias voltages.
The shaded regions correspond to the depth values where
the electric field has large systematic uncertainties.}
\label{elf.eps}
\end{figure}

\indent\
Using empirical parameterization of the field dependence
of the electron mobility~\cite{arora} and~\cite{caughey}, 
one can derive the electric field as a function of the depth
\begin{equation}
{E(z) = E_{ce}} \left[ \left( \mu_{0e} \over \mu_{e} (z) \right)^{\gamma_{e}} - 1 \right] ^{1/\gamma_{e}}, 
\label{el_field_formula}
\end{equation}
where $\mu_{e}$ is the measured electron mobility and 
$\mu_{0e}$ (low electric field electron mobility), $E_{ce}$  and $\gamma_{e}$ 
are known empirical parameters from~\cite{arora} and~\cite{caughey}.
The parametrized mobility agrees within 5\% with the measured mobilities~\cite{caughey}. 
The relative error of the electric filed  $r_E$ calculated using Eq.~\ref{el_field_formula}
is related to relative error of the mobility $r_\mu$ as
\begin{equation}
 r_E = r_\mu \frac{1}{1 - \left({\mu \over \mu_0}\right)^\gamma}.
\label{el_field_error_formula}
\end{equation}
For the expected mobility range the error of the electric field extraction method is between 5 and 15\%.
Fig.~\ref{elf.eps} shows the electric field obtained neglecting the electric field lines distortion close to the pixel implants.
The measurement is restricted to the depth range 17 \um$<z<268$ \um\
for the  reasons explained in section~\ref{Charge_collection_and_the_Lorentz_deflection}.
The errors shown in Fig.~\ref{elf.eps} are attributed to statistical fluctuation
of the collected charge and do not include the error of the electric field extraction method, 
which is below 15\%.
For the non-irradiated sensor the electric field is close to the classical linear field of an abrupt p-n junction.
For the heavily irradiated sensors the electric field has a double peak with a distinct minimum close to the middle of the bulk.
The origin of the double-peak electric field is qualitatively described in~\cite{Eremin_two_peak}.  
A two-trap model producing a doubly-peaked electric field was implemented in a detailed detector simulation 
and the simulated charge collection was found to be in 
good agreement with the measurements~\cite{mswartz_simulatioin,vchiochia_ieee04}.
By integrating the electric field over the depth one can determine the potential
drop across the silicon bulk.
The potential drop agrees with the applied bias voltage within 15\% for all sensors.

\subsection{Cross-check of the measured electric field}
In order to check the measured electric field a simulation of the
signal induced in the pixels was performed. 
The particle crosses the silicon sensor with an angle $\alpha=15$\Deg\ (see Fig.\ref{grazingangle_como.eps}) 
and the energy loss is assumed to be uniformly distributed. 
Neither energy loss fluctuation nor charge diffusion was taken into account.
In this simulation the electric field lines are assumed to be 
perpendicular to the silicon sensor planes and the electric field 
value as function of the depth was taken from the measurement shown in Fig. \ref{elf.eps}.
The time dependent induced current is calculated using the Shockley-Ramo theorem~\cite{shockley-ramo}
\begin{equation}
\begin{array}{lll}
i(t) & = & Q_h(t) \vec G[z(t)] \cdot \vec{v}_h[z(t)] +\\
 &  & Q_e(t) \vec G[z(t)] \cdot \vec{v}_e[z(t)], 
\end{array}
\end{equation}
where $Q_h$ and $Q_e$ are the holes and electrons charge values deposited by the particle energy loss, respectively, 
$\vec G$ is the weighting field,  $\vec{v}_h $  and $\vec{v}_e $ the holes and electrons drift velocities, respectively.

\begin{figure}[htb]
\begin{center}\epsfig{file=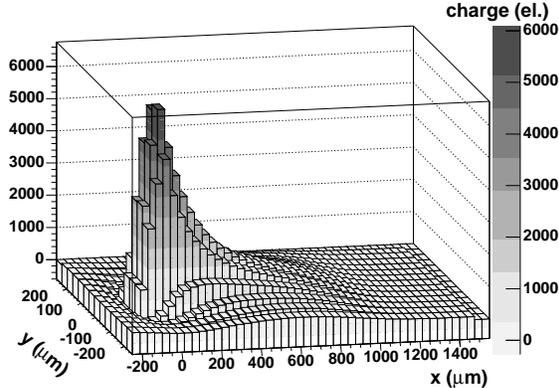,width=0.3\linewidth,bb=220 110 375 346, angle=0,silent=}\end{center}
\caption{Simulated signal induced by holes.}
\label{sim_cl_h_mag.eps}
\end{figure}

\begin{figure}[htb]
\begin{center}\epsfig{file=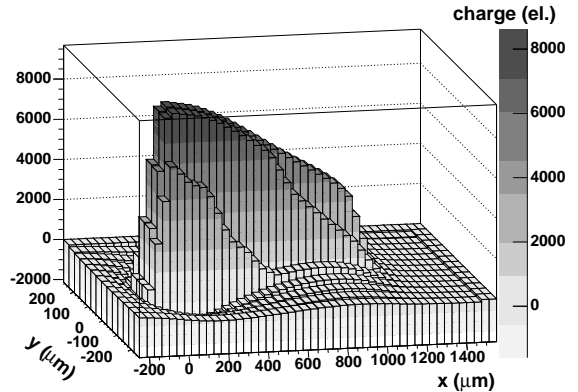,width=0.3\linewidth,bb=220 110 375 346,angle=0,silent=}\end{center}
\caption{Simulated total induced signal along the particle track projection with a magnetic field of $(3\cos(15\Deg))$ T.}
\label{sim_cl_mag.eps}
\end{figure}

\begin{figure}[htb]
\begin{center}\epsfig{file=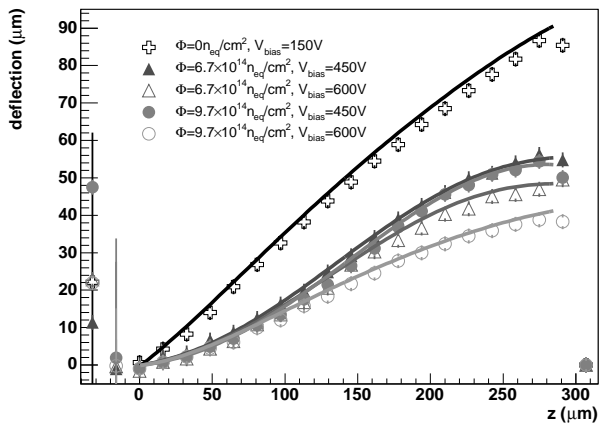,width=\linewidth,bb=20 90 490 346,angle=0,silent=}\end{center}
\caption{Measured (markers) and simulated (solid lines) deflection as a function of depth for different 
fluences and bias voltages.}
\label{la_comp.eps}
\end{figure}

\begin{figure}[htb]
\begin{center}\epsfig{file=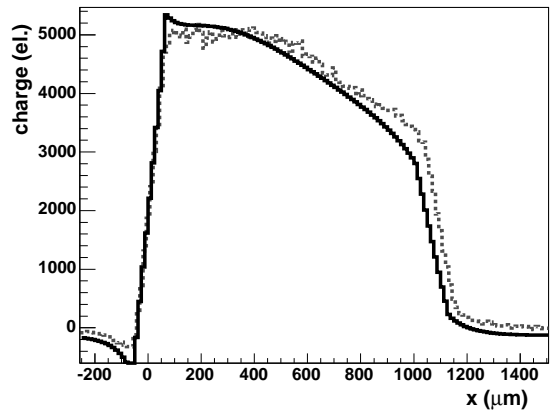,width=0.9\linewidth,bb=50 90 480 346,angle=0,silent=}\end{center}
\caption{Charge collected along the particle track for a sensor irradiated at 6.7\E{14}\Neq.
The solid line is the simulated charge, dashed line is the measured one.}
\label{cl_profiles.eps}
\end{figure}

The drift velocity is calculated using the measured electric field 
and the electrons and holes are trapped during the drifting time according to the exponential law
\begin{equation}
\begin{array}{ll}
Q_h(t) = Q_{0h} e^{-t/\tau_{h}}, &\:\: Q_e(t) = Q_{0e} e^{-t/\tau_{e}},
\end{array}
\end{equation}
where the fluence dependent trapping probabilities ${\tau_{h}}^{-1}$ 
and ${\tau_{e}}^{-1}$
are calculated assuming a linear dependence on irradiation fluence.
The proportionality coefficients are 4.2\E{-16} cm$^2/$ns for electrons
and 6.1\E{-16} cm$^2/$ns for holes respectively~\cite{Kramberger}.
For the non-irradiated sensors the trapping probability is set to zero,
as the collection time is in the order of few nanoseconds.
%
\begin{figure}[htb]
\begin{center}\epsfig{file=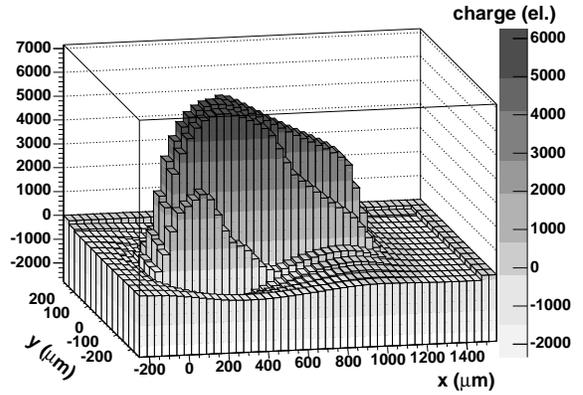,width=0.3\linewidth,bb=220 110 375 346,angle=0,silent=}\end{center}
\caption{Simulated signal induced by electrons.}
\label{sim_cl_e_mag.eps}
\end{figure}
The induced signal was calculated separately for holes (see Fig. \ref{sim_cl_h_mag.eps}) and
electrons (see Fig. \ref{sim_cl_e_mag.eps}) taking into account the Lorentz force.
The total induced signal is shown in Fig.~\ref{sim_cl_mag.eps}.
The contribution from holes is significant only at the region close to the pixel implant 
while the total induced current is dominated by electrons.
Figure \ref{la_comp.eps} shows the measured and simulated deflection as function of depth.
The charge deflection predicted by the simulation reproduces the measurements well.
The simulation was performed also without magnetic field to compare the charge collection
efficiency with the measurement.
The simulation reproduces the measured values very well (see Fig. \ref{cl_profiles.eps}).
The small discrepancies are due to charge diffusion, 
energy deposit fluctuation, electric field distortion between the implants
which were not implemented in the simulation.

\section{Summary}
A new method to extract the electric field in irradiated silicon pixel sensors is proposed 
and validated with a simulation.
The method is based on a precise measurement of the Lorentz deflection as a function of depth in the silicon sensor bulk. 
The extracted electric field is used in a sensor simulation which 
reproduces very well both the charge collection and the Lorentz deflection.\\ 
The method uses the electric field dependency on the mobility. Therefore, the precision 
of the method is limited by the precision of this dependence. 
However the empirical dependence agrees with the measurements very well~\cite{caughey}
and the error on the electric field is estimated to be less than 15\%. \\
The Lorentz angle is used for mobility calculation
and the error of the Lorentz angle measurements is dominated by the statistical 
fluctuations of the collected charge. 
The influence of the irradiation on the mobility and the Hall factor can be neglected
for fluences up to 10$^{15}$\Neq\ because of the lattice
scattering dominates over the scattering on the irradiation induced defects.

\section*{Acknowledgments}
  We gratefully acknowledge Silvan Streuli from ETH Zurich and Fredy Glaus from PSI for
their immense effort on the bump bonding of the pixel sensors.
We would like to thank Maurice Glaser and Michael Moll from CERN for 
carrying out the irradiation, Kurt B\"osiger from the Z\"urich workshop for the mechanical
construction, Gy\"orgy Bencze and Pascal Petiot from CERN for the H2 beam line support 
and finally the whole CERN-SPS team. 

\end{document}